\documentclass[pra,twocolumn,groupedaddress,showpacs,nofootinbib]{revtex4}
\newcommand{\beq}{\begin{equation}}
\newcommand{\eeq}{\end{equation}}
\newcommand{\beqa}{\begin{eqnarray}}
\newcommand{\eeqa}{\end{eqnarray}}
\usepackage{amsfonts}
\usepackage{graphicx}
\usepackage{amsmath}
\usepackage{bm}
\def\ra{\rangle}
\def\la{\langle}

\begin{document}

%\preprint{}

\title{Transient energy excitation in shortcuts to adiabaticity for the time dependent harmonic oscillator}

\author{Xi Chen$^{1,2}$}
\author{J. G. Muga$^{1,3}$}
%\author{....}
%\thanks{}
\affiliation{
   $^{1}$ Departamento de Qu\'{\i}mica-F\'{\i}sica,
UPV-EHU, Apartado 644, 48080 Bilbao, Spain}
\affiliation{
   $^{2}$ Department of Physics, Shanghai University,
200444 Shanghai, P. R. China}
\affiliation{
   $^{3}$ Max Planck Institute for the Physics of Complex Systems,
   N\"othnitzer Str. 38, 01187 Dresden, Germany}


%\date{\today}

\begin{abstract}
There is recently a surge of interest to cut down the time it takes to
change the state of a quantum system adiabatically.
We study for the time-dependent harmonic oscillator the transient energy excitation in speed-up processes
designed to reproduce the initial populations at some predetermined final frequency and time,
providing lower bounds
and examples. Implications for the limits imposed to the
process times and for the principle of unattainability of the absolute zero, in a single expansion or in quantum refrigerator cycles,
are drawn.
%in the form of a vanishing cooling rate when the cold temperature bath approaches zero. Otto cycles, are drawn.
% direct principle of zero
%and on the scaling of cooling rates that quantify the third law of thermodynamics in quantum Otto cycles
%are drawn.
\end{abstract}

 %02.30.Xx}
%\keywords{}

\maketitle
\section{Introduction}
Adiabatic processes in quantum systems are
frequently useful to drive or prepare states in a robust and controllable manner, and have also been proposed to solve complicated computational problems,
but they are, by definition, slow. (The definition of ``adiabatic process'' here is the usual one in quantum mechanics, namely, a slow change of Hamiltonian parameters keeping the populations of the instantaneous eigenstates constant all along.)
Thus a natural objective is to cut down the time to arrive at the
same final state, possibly up to phase factors, in other words, to find
``shortcuts to adiabaticity'', by designing optimal adiabatic pathways, or by admitting that the populations may not be preserved at intermediate times.
%and there is recently a surge of interest to cut down the time it takes to
%change the state of a quantum system adiabatically.
Several works have recently proposed different
ways to achieve this goal for general or specific cases
\cite{ion,David08,Berry09,Muga09,MN10,Muga10,Ch10,Ch10b,Calarco09,MN10b}.
One of the early applications considered has been particle transport without vibrational heating \cite{ion,David08,Calarco09,MN10,MN10b}. Another important case is frictionless harmonic trap compressions or expansions for state preparation \cite{Salamon09,Muga09,MN10,Ch10,Muga10,Ch10b}, which were first addressed with ``bang-bang'' (piecewise constant frequency) methods \cite{Salamon09};
other route is to design by inverse engineering techniques a time dependent frequency for which the expanding modes
associated with Lewis-Riesenfeld invariants \cite{LR} take the state from the initial to the final potential configuration without transitions \cite{Ch10,Muga10}; this has been extended to the Gross-Pitaevskii equation with a variational ansatz \cite{Muga09},
and has been also implemented experimentally
to decompress $^{87}$Rb cold atoms
in a harmonic magnetic trap \cite{Nice}.
In the same vein, Berry has provided an algorithm to construct a Hamiltonian $\widetilde{H}(t)$ for which the adiabatic approximation for the state evolution under a time-dependent reference Hamiltonian
$H(t)$ becomes the exact dynamics with $\widetilde{H}(t)$. This algorithm has been applied to spins
in magnetic fields \cite{Berry09}, harmonic oscillators \cite{Muga10}, or
to speed up adiabatic state-preparation methods such as Rapid Adiabatic Passage (RAP), Stimulated Rapid Adiabatic Passage (STIRAP) and its variants \cite{Ch10b}.
Also, Masuda and Nakamura have adapted for adiabatic processes \cite{MN10} a (``fast-forward'') scaling technique to speed up the state dynamics \cite{MN08},
with application examples to particle transport or time dependent harmonic potentials \cite{MN10}, and spins or charged particles
in electro-magnetic fields \cite{MN10b}. Finally, optimal control theory has also been used for nonadiabatic cooling under imposed costs \cite{Rabitz98,Salamon09}; and Lyapunov control methods have been proposed to speed up quantum
adiabatic computing without information of the Hamiltonian eigenstates \cite{Wang10}.

In this paper we shall examine the energy ``cost'' of such processes; more precisely, their transient excitation energies. Our central study case
is the expansion (or compression) of a harmonic oscillator, a basic model for many
operations in any cold atoms laboratory \cite{Ch10}.
%, although some of the results are generally applicable, and other relevant examples concerning
%discrete systems (such as accelerated versions of Rapid Adiabatic Passage) are also discussed briefly.
Intuitively, one expects the transient system energy and the time of the process to be ``conjugate'',
i.e., an increase of the former when decreasing the later,
but the details of this relation, and the role played by other parameters defining the process
(such as initial and final frequencies) have to be clarified
% \cite{b1,b2}.
both for fundamental reasons and for the applications. In particular, the energy excitation will set limits to the possible speed-up. In a trap which is harmonic near the ground state but not for higher energies, large transient energies will imply perturbing effects of anharmonicities and thus undesired
excitations of the final state, or even atom loss.
% In a SHAPE process high energies would make the
% process comparable in efficiency to simpler procedures.
The transient excitation energy has also implications for quantifying the
principle of unnatainability of zero temperature, first
enunciated by Nernst \cite{Nernst}.
Fowler and Guggenheim \cite{FG} formulate it as follows:
``{\em{It is impossible by any procedure no matter how idealized to reduce the temperature of any system to the absolute zero in a finite number of operations.}}''
They identify it with the third law of thermodynamics although this is sometimes disputed.
More recently, Kosloff and coworkers  \cite{K2000,Salamon09,Kosloff-EPL} have restated the unattainability principle as the vanishing
of the cooling rate in quantum refrigerators  when the temperature of the cold bath approaches zero, and quantify it by the scaling law relating
cooling rate and cold bath temperature.
We shall examine the consequences of the transient energy excitation
on the unattainability principle at two levels, namely,
for a single, isolated expansion, and considering  the expansion as one of the branches of a quantum refrigerator cycle.

When describing these cycles and indeed in many intersection areas between
quantum mechanics and thermodynamics one finds the need to use the word ``adiabatic'' in two
different ways: the thermodynamical one (meaning that there is no heat transfer between system and environment) and the quantum one. Many authors have pointed out this
duality as an unfortunate source of confusion.
It may prove useful to distinguish  them and avoid ambiguities and the hassle of detailed explanations with a shorthand notation. Following the example of Dirac's $q$-number versus $c$-number distinction, we propose to
refer to a process as ``$t$-adiabatic'' if it is thermodynamically adiabatic,
and as ``$q$-adiabatic'' if it is a quantum mechanically adiabatic (i.e. slow) process.

%
%
%The paper is organized as follows.
%We begin with the shortcut to adiabaticity in Harmonic oscillator
%in Sec. \ref{model}, and the time-averaged energy and the uncertainty of energy are also given.
%In Sec. \ref{A-A}, we discuss the bound for time-averaged uncertainty
%of energy obtained from the well-known
%Anandan-Aharonov relation, which is too optimistic and not dependent of finial frequency.
%To overcome these difficulties, we adopt in Sec. \ref{bound} the variational principle to
%achieve the bound by quasi-optimal trajectory and minimize the time-averaged energy.
%In addition, the comparisons with the ``Bang" or ``Bang-Bang" method are also made in
%Sec. \ref{bang-bang}. Finally, we will summarize the paper in Sec. \ref{conclusion}.
%
%
%
%
%
\section{Bang-bang methods}
\label{bang-bang}
%
%
%
%
%
%\subsection{Bang bang}
%
%
%
%
%
%
The Hamiltonian for a particle with mass $m$ in a time-dependent harmonic oscillator is given by
\beq
H=\hat{p}^2/2m+m\omega^2(t)\hat{q}^2/2.
\label{H}
\eeq
Let us assume an expansion (compressions are treated similarly) with initial angular frequency $\omega_0\equiv\omega(0)$ at time $t=0$ and final frequency $\omega_f\equiv\omega(t_f)<\omega(0)$
at time $t_f$.

In the ``bang-bang'' approach the frequency is shaped  as a stepwise
constant function of time, choosing the step values and durations
so as to preserve the initial state populations in the final configuration. {}For {\it real} trap intermediate frequencies this requires a minimal total expansion time \cite{Salamon09}
\beq
\label{bb}
t_f>\frac{\sqrt{1-\omega_f/\omega_0}}{\sqrt{\omega_f\omega_0}}.
\eeq
%
%If, in addition, the frequency is constrained to lie within the
%interval $[\omega_i,\omega_f]$,
%then
%
%\beq
%t_f>\frac{1}{2}\left(\frac{1}{\omega_i}+\frac{1}{\omega_f}\right)\arccos\left(\frac{\omega_i^2+\omega_f^2}{(\omega_i+\omega_f)^2}\right).
%\eeq
%
The limit can be realized by only three jumps, i.e.,
two real intermediate frequencies
(it cannot be improved
by using more intermediate frequencies),
but one of the
intermediate frequencies should be infinite.

%This is regarded as an example for the finite-time
%third law of thermodynamics.
Up to a constant factor the main dependence in the bound (\ref{bb}) already appears in a simpler process that
reproduces for $(\omega_f,t_f)$ the initial populations
with just one intermediate frequency, the geometric average
$\omega_1=(\omega_0\omega_f)^{1/2}$,
and a total time
\beq
\label{b}
t_f=\frac{\pi}{2 \sqrt{\omega_0\omega_f}},
\eeq
%
%In this process with two jumps, the average energy is always $\overline{E}_0 = \frac{\hbar \omega_0/2 + \hbar \omega_f/2}{2}$.
which is a quarter of the corresponding period \cite{Vogels}.
For an initial $n$-th state of the oscillator the instantaneous mean energy during the transient period becomes
the arithmetic mean of the initial and final energies,
\beq
\label{bha}
\la H\ra= \hbar\left(n+\frac{1}{2}\right)\frac{\omega_0 + \omega_f}{2}.
\eeq
Eq. (\ref{b}) and the bound (\ref{bb}) are relevant because the $t$-adiabatic expansion is actually the speed bottleneck in quantum Otto refrigerator cycles which use particles in a harmonic oscillator as the ``working medium'' \cite{Kosloff-EPL,Salamon09}.
The cooling rate $R$ as the cold bath temperature $T_c$ approaches zero
is dominated by the expansion time and scales as $R\propto\omega_f/t_f$. Since $\omega_f\propto T_c$ as $T_c\to 0$, the dependence of $t_f$ on $\omega_f$  quantifies the unattainability principle.
In particular, $q$-adiabatic
%(here meaning slow as noted above)
expansions lead to $R\propto T_c^3$ scalings, in contrast with the $R\propto T_c^{3/2}$ scaling achieved with the times in Eqs. (\ref{bb})
or (\ref{b}) \cite{Kosloff-EPL,Salamon09}.

In \cite{Ch10} it was demonstrated, however, that the
minimal time in Eq. (\ref{bb}) can be beaten with  bang-bang methods and inverse engineering methods,
see also the next section,
by allowing for imaginary intermediate frequencies, i.e., transients
in which the harmonic oscillator becomes a parabolic repeller. It was pointed out \cite{Ch10}, that this new freedom leads to the absence, at least in principle,  of a lower bound for the expansion time, which could obviously affect the optimal scaling of cooling rates. We shall analyze the impact of these ultra-fast expansions on the third law in the following sections.
\section{Energy bounds for inverse-engineered time-dependent harmonic oscillators\label{inv}}
\subsection{Bound for time-averaged energy}
In this subsection, we will set a lower bound for the time-averaged energy in the
transitionless expansions and compressions of the time-dependent harmonic oscillator.
A shortcut to adiabaticity taking the $n$-th state of the initial trap to the final $n$-th state of the final trap up to phase factors is achieved \cite{Ch10} by designing the
frequency from the Ermakov equation
\beq
\label{Ermakov}
\ddot{b}+\omega^2(t) {b}=\frac{\omega_0^2}{b^3},
\eeq
where $b$ is an engineered scaling function
which satisfies the following boundary conditions
at $t=0$ and $t_f$,
\beqa
b(0)=1, ~ \dot{b}(0)=0, ~ \ddot{b}(0)=0,
\nonumber\\
b(t_f)=\gamma, \dot{b}(t_f)=0, \ddot{b}(t_f)=0.
\label{bc}
\eeqa
Here
$\gamma = \sqrt{\omega_0/\omega_f}$, and the single and double dots denote first and second derivatives with respect to time.
The simplest choice for interpolating $b(t)$  between $0$ and $t_f$ is a polynomial form,
$b(t)=6(\gamma-1) s^5-15(\gamma-1) s^4+ 10(\gamma-1) s^3+1 $,  where $s=t/t_f$.
In this manner the $n$-th stationary state of the initial oscillator will evolve according to the ``expanding mode''
\beqa
\label{single mode}
\Psi_n (t,x) &=& \left(\frac{m\omega_0}{\pi\hbar}\right)^{1/4}
\!\frac{e^{-i (n+1/2) \int_0^t dt'\, \frac{\omega_0}{b(t')^2}}}{(2^n n! b)^{1/2}}
\nonumber\\
&\times&e^{i \frac{m}{2\hbar}\left(\frac{\dot{b}}{b(t)} +
 \frac{i\omega_0}{b^2}\right)x^2}
{\cal{H}}_n\left[\left(\frac{m\omega_0}{\hbar}\right)^{1/2}\frac{x}{b}\right],
\eeqa
where ${\cal{H}}_n$ is a Hermite polynomial, and will become eventually, up to a phase, the
$n$-th eigenstate of the final trap at $t_f$.
At intermediate times $|\Psi_n\ra$ does not coincide in general with the instantaneous eigenvectors $|n\ra$ of $H(t)$, $H(t)|n(t)\ra=\epsilon_n(t)|n(t)\ra$.

For the $n$-th expanding mode, the instantaneous average energy $E_n(t) \equiv\langle \Psi_n|H (t)|\Psi_n \rangle$ is
\beq
\label{energy}
E_n(t)=\frac{(2n+1)\hbar}{4\omega_0}
\left[\dot{b}^2+\omega^2(t)b^2+\frac{\omega_0^2}{b^2}\right],
\eeq
which is in general different, except at initial and final times, from $\epsilon_n$.
The time average of $E_n$ is defined by
\beq
\label{average energy}
\overline{{E}_n} \equiv\frac{1}{t_f}\int^{t_f}_0 E_n(t)\, dt.
\eeq
To find a lower bound for $\overline{E_n}$
we substitute
Eq. (\ref{energy}) into Eq. (\ref{average energy}),
and integrate by parts making use of the boundary conditions (\ref{bc}),
%to eliminate the $\ddot{b}$ term,
%
\beq
\label{energy-integral}
\overline{{E}_n}= \frac{(2n+1)\hbar}{2\omega_0 t_f} \int^{t_f}_0
\left(\dot{b}^2+\frac{\omega_0^2}{b^2}\right) dt.
\eeq
The integrand has the form of the Lagrangian of a particle in an attractive inverse square potential, but
the minimization problem, i.e., finding an optimal function $b(t)$ subjected to
the boundary conditions (\ref{bc}) cannot be solved with an ordinary Euler-Lagrange equation since there are too many boundary conditions which affect not only $b$ but also $\dot{b}$ and $\ddot{b}$ at the edges of the time interval.
We can nevertheless find easily, using the Euler-Lagrange equation,
the quasi-optimal ``trajectory'' $b(t)$ that minimizes the integral
subjected only to the boundary values of $b$, that is, $b(0)=1$ and $b(t_f)=\gamma$. Since these two conditions define a broader set of functions than the ones satisfying (\ref{bc}), the quasi-optimal $b$  provides at least a lower bound for the time-averaged energy.
For the function
\beq
f (t, b, \dot{b}) =  \dot{b}^2 + \omega_0^2/b^2,
\eeq
the Euler-Lagrange differential equation
$f_b- \frac{d}{dt} f_{\dot{b}}=0$
is
\beq
\label{equation for optimal trajector}
b^3 \ddot{b}= -\omega^2_0.
\eeq
The solution satisfying $b(0)=1$ and $b(t_f)=\gamma$
is
%a quasi-optimal trajectory $b(t)$ that
%at least provides a bound for the time-averaged energy,
%
\beqa
\label{optimal trajector}
b (t)&=& \sqrt{(B^2-\omega^2_0 t^2_f) s^2 +  2 B s + 1},
\eeqa
where $B=-1 + \sqrt{\gamma^2 + \omega^2_0 t^2_f}$
and the positive root should be taken.
Substituting Eq. (\ref{optimal trajector}) into the integral (\ref{energy-integral}),
we finally obtain a lower bound for the
time-averaged energy,
\beqa
\label{minimal energy}
{\cal B}_n &=&  \frac{(2n+1)\hbar}{2 \omega_0 t_f^2} {\Big\{} (B^2-\omega^2_0 t^2_f)-2 \omega_0 t_f
\nonumber
\\&\times&{\Bigg [}
\mbox{arctanh}\left(\frac{B^2+B-\omega^2_0 t_f^2}{\omega_0 t_f} \right)
\nonumber\\
&-&
\mbox{arctanh}\left(\frac{B}{\omega_0 t_f} \right)
{\Bigg]}{\Bigg\}},
\eeqa
such that $\overline{E_n}\ge{\cal B}_n$.
When the final frequency $\omega_f$ is small enough to satisfy $t_f \ll 1/\sqrt{\omega_0 \omega_f}$, and $\gamma \gg 1$, the
lower bound has the following simple asymptotic form,
\beq
\label{asymptotic}
{\cal B}_n\approx \frac{(2n+1) \hbar}{2 \omega_f t^2_f}.
\eeq
Incidentally, ${\cal{B}}_n$ sets also a lower bound for the maximum of the instantaneous energy $E_n(t)$.
%------------------fig 1 begin-------------------
\begin{figure}[ht]
\begin{center}
\scalebox{0.35}[0.35]{\includegraphics{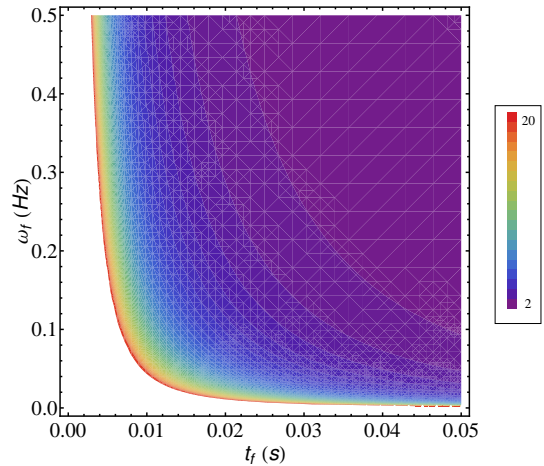}}
\end{center}
\caption{\label{fig.1} (Color online) Lower bound for time-averaged energy (in units of $E_0(0)=\hbar\omega_0/2$) as a function of $t_f$ and $\omega_f$, where $\omega_0= 2 \pi \times 250$ Hz.}
\end{figure}
%-----------------fig 1  end---------------------

\begin{figure}[ht]
\begin{center}
\scalebox{0.45}[0.45]{\includegraphics{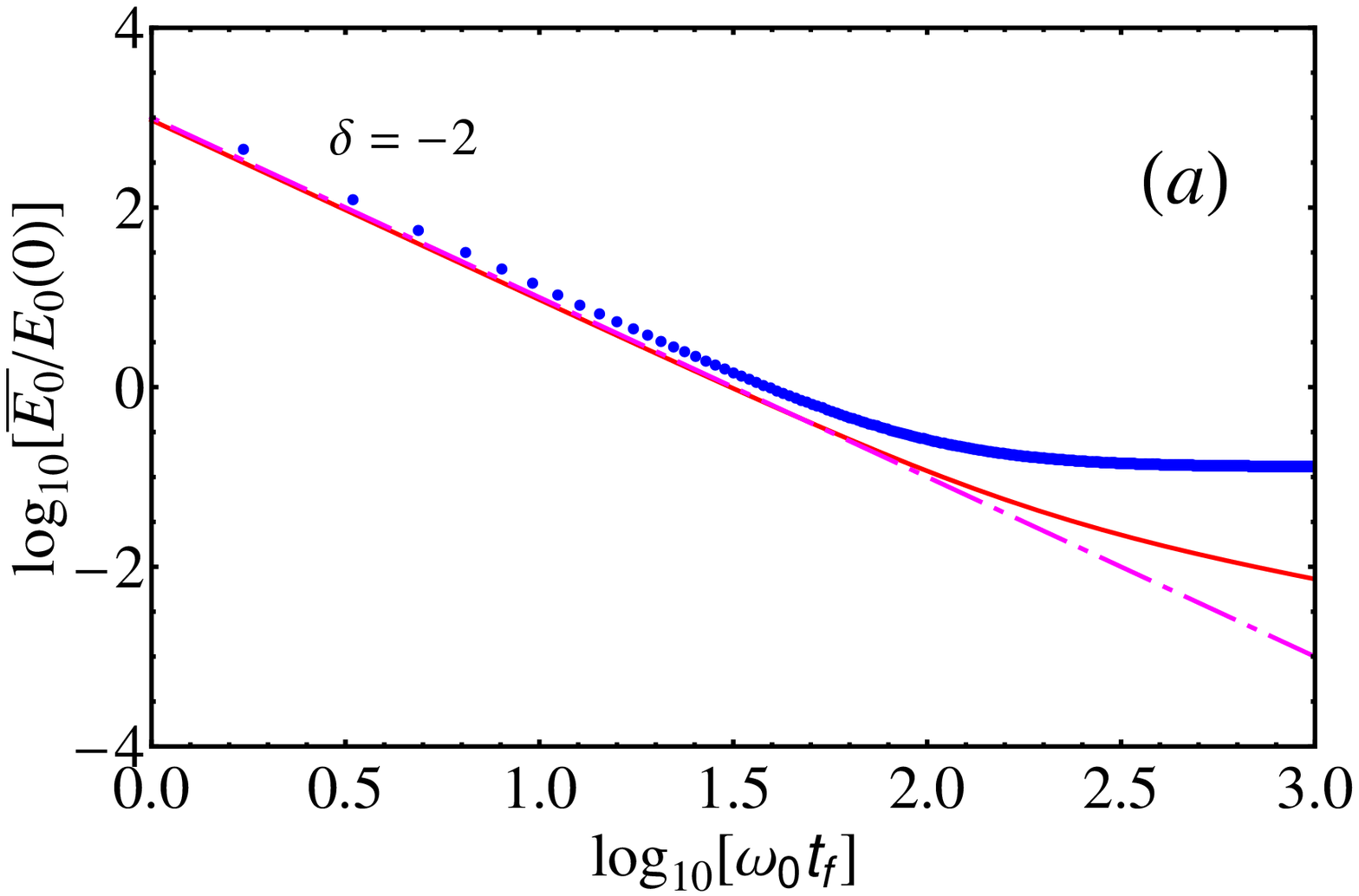}}
\scalebox{0.45}[0.45]{\includegraphics{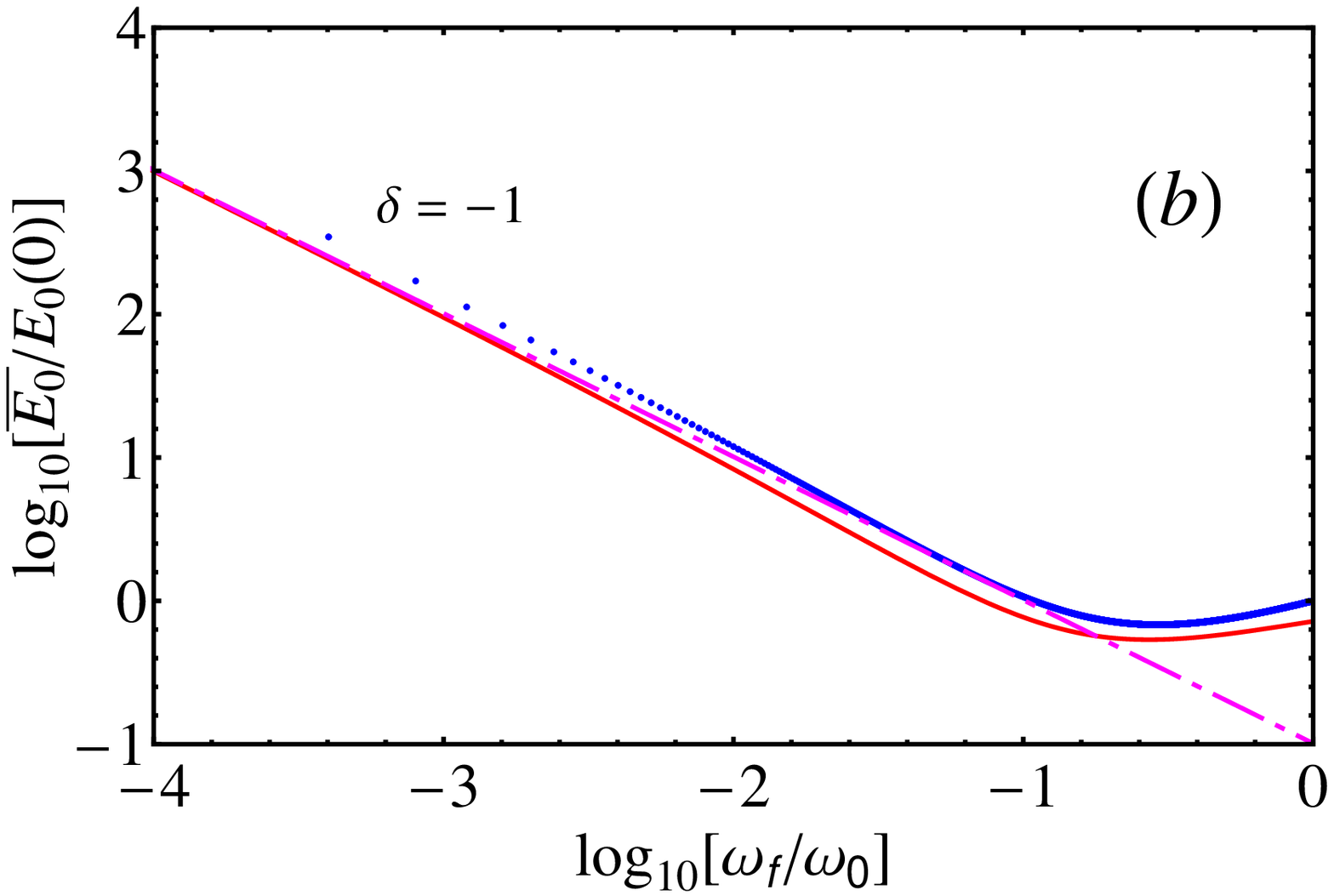}}
\end{center}
\caption{\label{fig.2} (Color online) Dependence of the time-average energy for the ground state $\overline{{E}_0}/E_0(0)$
on the (a) short time
$t_f$ ($\omega_f = 2 \pi \times 0.25$ {Hz}) and (b) final frequency $\omega_f$ ($t_f = 2$ {ms}). In both cases $\omega_0 = 2 \pi \times 250$ {Hz}.
Bound given by Eq. (\ref{minimal energy}) (solid red line); asymptotic expression Eq. (\ref{asymptotic}) (dot-dashed magenta line), and time-average energy for polynomial $b$ (dotted blue line). The $\delta$'s  are the asymptotic exponents of $t_f$ and $\omega_f$ respectively as they go to zero.}
\end{figure}

In Fig. \ref{fig.1} we plot ${\cal B}_0$ as a function of $t_f$ and $\omega_f$
for $\omega_0= 2 \pi \times 250$ Hz, which will be also the
initial frequency in the following examples.
The important point is that the transient energy increases not only with decreasing final time $t_f$ but also with decreasing final frequency $\omega_f$. Figure \ref{fig.2} shows the exponents of the scaling
for the bound (\ref{minimal energy}), its asymptotic form (\ref{asymptotic}),  and the time-averaged energy for a polynomial $b$. In all cases $\overline{{E}_0}$, or ${\cal B}_0 \propto 1/(\omega_f t^{2}_f)$ asymptotically.

%In this case,
%
%\beq
%\label{energy bound}
%\overline{{E}}_n \geq  \frac{(2n+1) \hbar}{2 \omega_f %t^2_f},
%\eeq
%
A consequence of Eq. (\ref{asymptotic}) is
\beq\label{tfb}
t_f \geq \sqrt{\frac{(2n+1) \hbar}{2 \omega_f \overline{E_n}}}.
\eeq
%
%for the fixed average energy $\overline{{E}}_n$.
%For fixed $\overline{E_n}$, this is the same scaling
%with respect to $\omega_f$ found by real-frequency bang-bang methods, and it also holds for the time-averaged energy for a polynomial $b$.

%
%------------------fig 2 begin-------------------
%-----------------fig 2  end---------------------
%
%This is the same scaling in the asymptotic limit
%$t_f \ll 1/\sqrt{\omega_0 \omega_f}$ as the low energy bound.
%In addition, the bound for the time-averaged energy (\ref{energy bound}) %also gives the same scaling with respect to $\omega_f$ as in bang-bang  %methods, if
%the energy is a finite resource, as it is.
The interest of Eq. (\ref{tfb}) compared to Eq. (\ref{bb})
is that in principle, for fixed $\omega_0$, it is possible to beat
the bang-bang minimal time, but the price is an increase
in the transient energy. In practice this energy cannot be arbitrarily large, if only because there are no perfect harmonic oscillators.
In particular, if we consider in Eq. (\ref{tfb}) that
$\overline{E_n}$ is limited by some maximal value, because of anharmonicities or a finite
trap depth, the obtained scaling is fundamentally the same as for bang-bang methods, and leads to a cooling rate $R\propto T_c^{3/2}$ in an inverse  quantum Otto cycle, although an opportunity is offered to improve the proportionality factor by increasing
the allowed $\overline{E_n}$.
%
%
%\subsection{Unattainability of zero temperature}
%
%

Independently of the participation of the harmonic trap expansion as a
branch in a refrigerator cycle, we may see rather directly the effect of the above analysis on a single expansion by assuming that the initial and final states are described by canonical density operators characterized by temperatures $T_0$ and $T_f$,
related by $T_f=(\omega_f/\omega_0)T_0$ for a population-preserving process.
%(This is the principle of ``adiabatic cooling'' when %the expansion is done slowly).
%(This is so because, even though the populations do not %change, the potential, energy levels and instantaneous %states do change.)
Within the idealized but specific context of a pure
parabolic potential expansion, the unattainability of a zero
temperature can be reformulated
microscopically as follows:
%in terms of energy rather than in terms of time, since in this setting  there is no lower limit to the process time, but
The transient excitation energy becomes infinite for any
population-preserving and finite-time process if the final temperature is zero (which requires $\omega_f=0$). This excitation energy has to be provided by an external device, so the absence of a lower process time limit should not lead us astray, since there remains a fundamental obstruction to reach $T_f=0$ in a finite time, in the form of the need for a source of infinite power.

\subsection{Minimization of time-averaged energy}
\begin{figure}[ht]
\begin{center}
\scalebox{0.50}[0.50]{\includegraphics{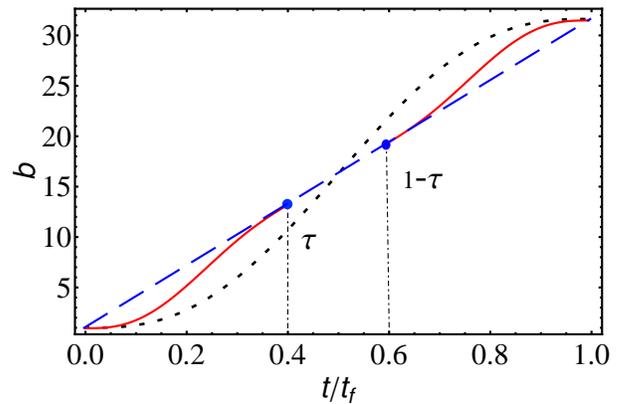}}
\end{center}
\caption{\label{fig.3} Example of hybrid $b$ combining the quasi-optimal trajectory in the
central segment $[\tau, 1-\tau]$ and ``cap polynomials"
with the right boundary conditions,
where $\omega_0 = 2 \pi \times 250$ Hz, $\omega_f = 2 \pi \times 0.25$  Hz,
$t_f = 2$ ms, $\tau=0.4$ (solid red line),
quasi-optimal square-root $b$ of Eq. (\ref{optimal trajector}) (dotted black line), and
polynomial trajectory (blue dashed line).}
\end{figure}

In order to minimize the time-average energy  and approach the lower bound, we can use the quasi-optimal  $b(t)$, Eq. (\ref{optimal trajector}),  in a
central time segment $[\tau, 1-\tau]$, and match it at the extremes with two ``cap polynomials", each of them satisfying three of the  boundary conditions at Eq. (\ref{bc}) (at $t=0$ or $t_f$), plus
three boundary conditions  for $b$, $\dot{b}$ and $\ddot{b}$ at the matching point.
The idea is illustrated in Fig. \ref{fig.3}.
%, where  $\omega_0 = 2 \pi \times 250 \mbox{Hz}$, $\omega_f = 2 \pi \times %0.25  \mbox{Hz}$,
%$t_f = 2 \mbox{ms}$, $\tau=0.4$ (solid red line),
%quasi-optimal square-root trajectory (dotted black line), and
%polynomial trajectory (blue dashed line).

%---------------------- fig3. begin-------------------------
%-------------------- fig3. end------------------------------

The resulting hybrid $b$ takes the form
\begin{eqnarray}
b = \left\{
\begin{array}{ll}
~~~~~~~ \sum_{j=0}^5 c_j s^j &~~ (0 \leq s \leq \tau)
\\
\sqrt{(B^2-\omega^2_0 t^2_f) s^2 +  2 B s + 1} & (\tau \leq s \leq 1-\tau)
\\
~~~~~~~ \sum_{j=0}^5 d_j s^j & (1- \tau \leq s \leq 1)
\end{array}
\right.
\label{hybrid}
\end{eqnarray}
where the coefficients $\{c_j\}$ and $\{d_j\}$ have lengthy expressions but are easily obtained from the matching conditions so we omit their explicit forms here.
%\beqa
%b(t) &=& \nonumber 1  + \frac{s^4}{\tau^4}\left[15 - 15 b(\tau) +7 \tau %\dot{b}(\tau) - \tau^2 \ddot{b}(\tau) \right]  \\  \nonumber
%&-& \frac{s^3}{2 \tau^3} \left[20 -20 b(\tau) + 8 \tau \dot{b}(\tau) - %\tau^2 \ddot{b} (\tau) \right]
%\\ &-& \frac{s^5}{2 \tau^5}\left[12 - 12 b(\tau) + 6 \tau \dot{b}(\tau) - %\tau^2 \ddot{b}(\tau)\right],
%\eeqa
%and
%\beqa
%b(t) &=& \nonumber \frac{(s-1+\tau)^3 }{\tau^5} \left[6 +6 s^2 + \tau (3 + %\tau) - 3 s (4 + \tau) \right]  \gamma \\  \nonumber
%&-& \frac{(s-1)^3}{\tau^5} \left[6 (s-1)^2 - 15 \tau (s-1) + 10 \tau^2 %\right] b(1-\tau)
%\\ \nonumber &+& \frac{(s-1)^3 (s-1+\tau) }{\tau^4}  (3- 3s -4 \tau ) %\dot{b}(1-\tau)
%\\ &-& \frac{(s-1)^3 (s-1+\tau)^2 }{2 \tau^3}  \ddot{b}(1-\tau).
%\eeqa

\begin{figure}[ht]
\begin{center}
\scalebox{0.50}[0.50]{\includegraphics{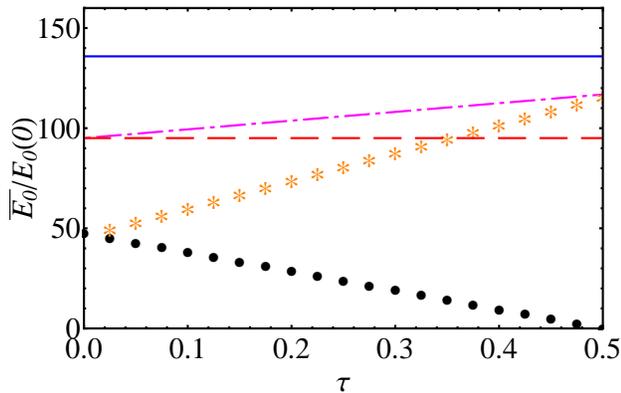}}
\end{center}
\caption{\label{fig.4} (Color online) Time averaged energy versus $\tau$ for polynomial $b$ (solid blue line), bound in Eq. (\ref{minimal energy}) (red dashed line), and
hybrid $b$ (dot-dashed magenta line). The contribution from the ``cap polynomials" (orange line with ``$\ast$") and
the central segment (black line with ``$\bullet$")
are also depicted.
Parameters are the same as in Fig. 3.}
% and
%without ``cap polynomials'' (dotted black lines).}
\end{figure}

Figure \ref{fig.4} demonstrates that
this hybrid $b$ can indeed minimize the time-averaged energy by making $\tau$ smaller and smaller,
approaching the lower energy bound as $\tau \rightarrow 0$. %However, it is seen from Fig. \ref{fig.3} that the polynomial caps contribute with a non-vanishing energy residue. As a matter of fact, we just obtained the above quasi-minimal energy by minimizing the integral (\ref{energy-integral}), instead of (\ref{average energy}).
%In other word, it suggests that the lower energy cost for the optimal square-root trajectory,
%depicted by dotted line in Fig. \ref{fig.3}, can't be achieved in the shortcut adiabaticity, due to the abundant boundary conditions.
A detailed calculation shows that the contribution from the caps does not vanish as $\tau\to 0$, so the value of the time-average bound is reached at the price of a singular instantaneous energy, see Fig. \ref{fig.4}.
%---------------------- fig4. begin-------------------------
%---------------------- fig4. end-------------------------

%
\section{Energy variance and Anandan-Aharonov relation}
\label{A-A}
We shall now discuss the impact of the shortcuts to adiabaticity on the standard deviation of the energy
$\Delta H\equiv(\la H^2\ra-\la H\ra^2)^{1/2}$. This is important because, a small averaged energy could in principle be spoiled by a large standard deviation.
%Fleming \cite{Fleming},
Anandan and Aharonov \cite{AA}
% and Vaidman \cite{Vaidman}
found a relation between the time average
of the standard deviation of the energy and the time of a process connecting two given states, irrespective of the Hamiltonian used to connect them. The so called ``Anandan-Aharonov" (AA) relation
provides a lower bound for the average uncertainty of the energy, which is extensively used to minimize the time $t_f$ required for the evolution between the two orthogonal quantum states.
Based on the Fubini-Study metric, the following distance may be defined,
\beq
\label{metric}
S= 2 \int^{t_f}_0 \frac{\Delta H (t)}{\hbar} dt \geq S_0,
\eeq
where the minimal value $S_0 = 2 \arccos(|\langle \Psi (t=0)| \Psi (t=t_f)\rangle|)$ corresponds
to the ``geodesic'' \cite{AAPati}. For orthogonal states $S_0=\pi$, and
\beq
\label{A-Arel}
\overline{\Delta H}\,   t_f \geq  \frac{h}{4},
\eeq
where
\beq
\overline{{\Delta H}}= \frac{\int^{t_f}_0 \Delta H (t) dt}{t_f},
\eeq
but more generally, for arbitrary (possibly  non-orthogonal) initial and final states, the AA relation is
\beq
\overline{\Delta H}\, t_f \geq \frac{h S_0}{4 \pi}.
\eeq
%
%------------------fig 5 begin-------------------
\begin{figure}[ht]
\begin{center}
\scalebox{0.45}[0.45]{\includegraphics{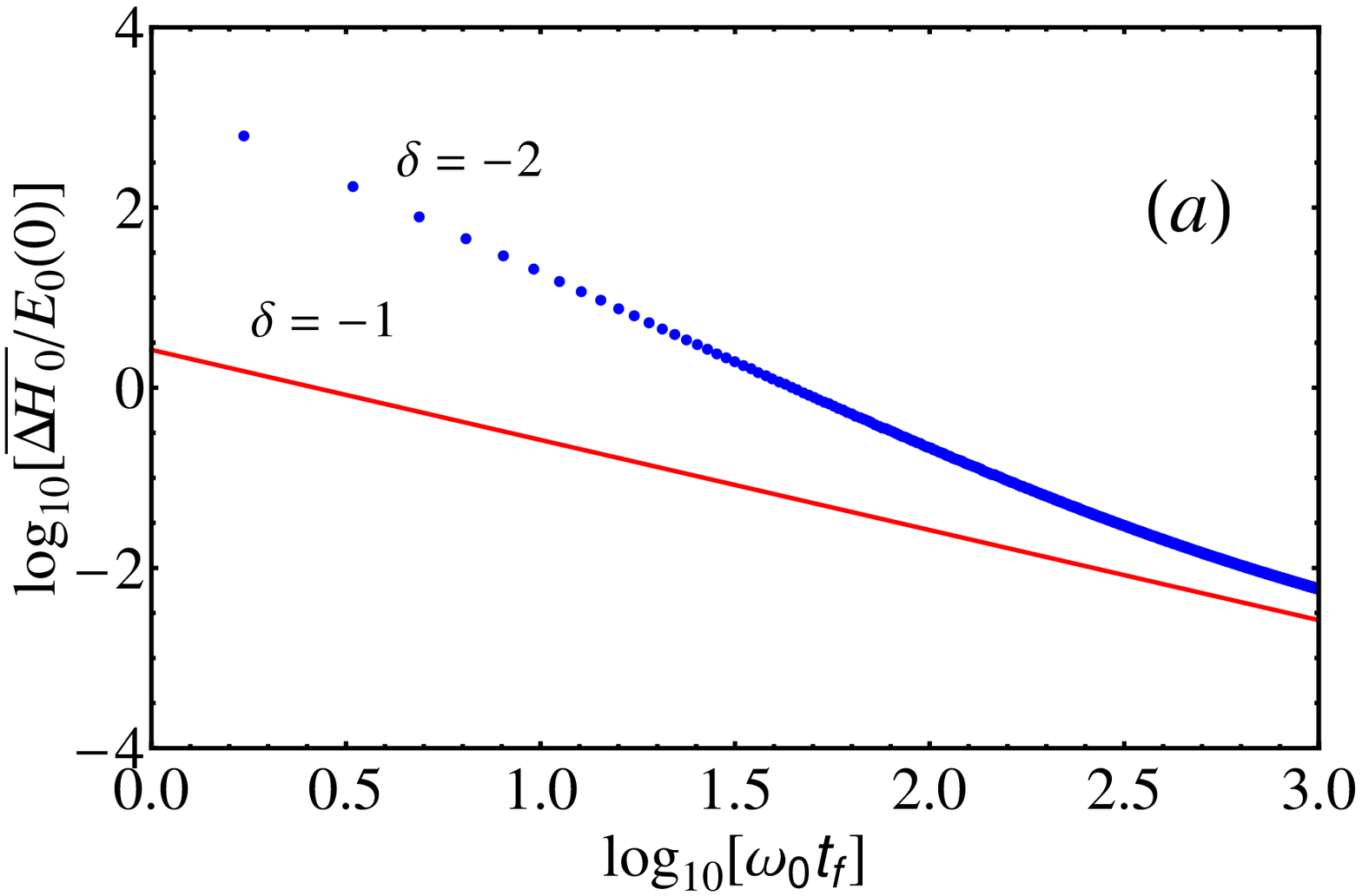}}
\scalebox{0.45}[0.45]{\includegraphics{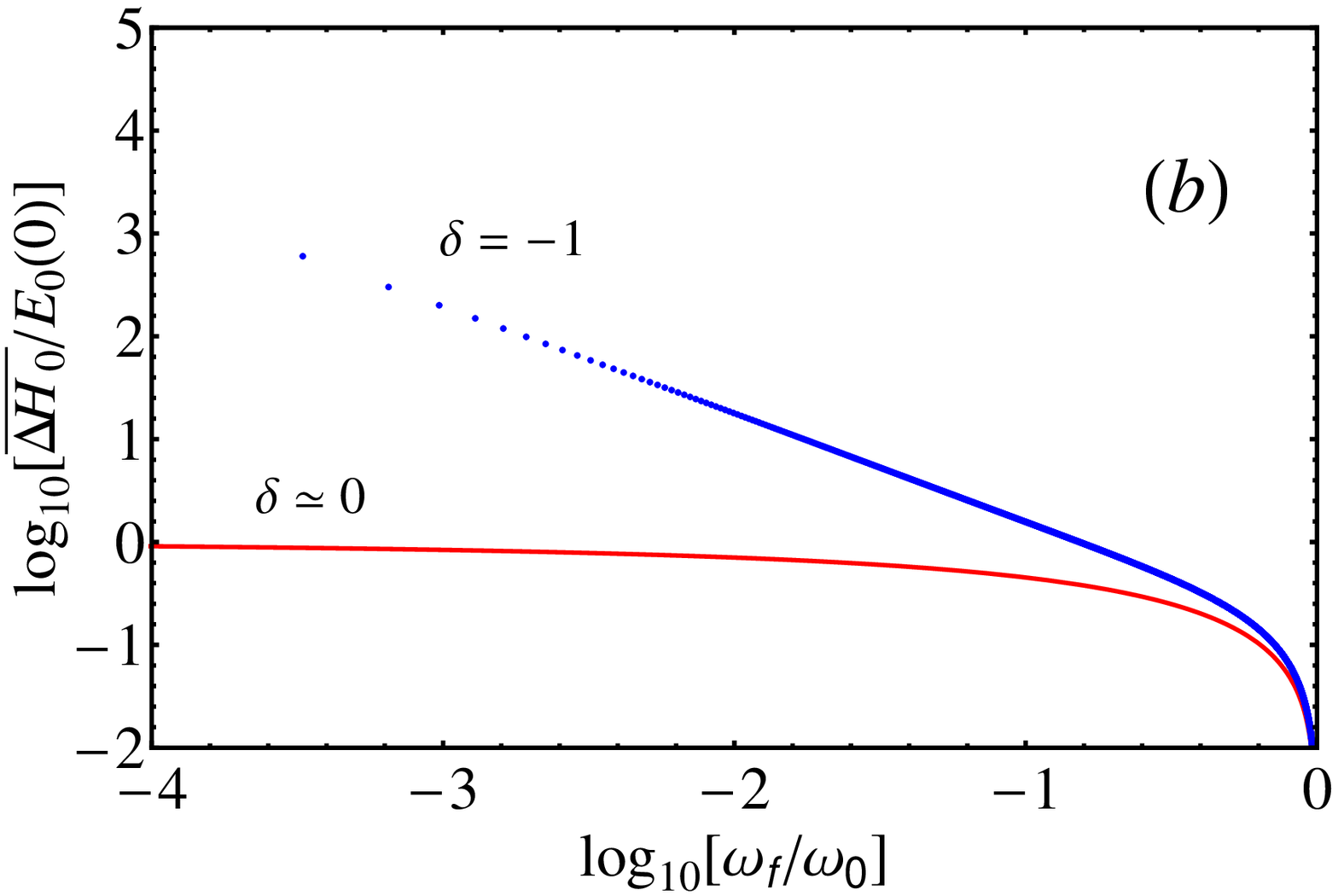}}
\end{center}
\caption{\label{fig.5} (Color online) Dependence of the time-averaged uncertainty of energy $\overline{\Delta H_0}$
on the (a) short time
$t_f$ and (b) final frequency $\omega_f$, where the parameters are the same as Fig. \ref{fig.2},
polynomial trajectory (dotted blue line),
AA relation (solid red line)
.}
\end{figure}
%-----------------fig 5 end---------------------
This may be applied to the harmonic oscillator and in
particular to any process taking an $n$-th initial eigenstate to an $n$-th final eigenstate, up to phase factors.
For the ground state  we find,
see (\ref{single mode}),
\beq
\cos^2\left(\frac{S_0}{2}\right) \equiv |\langle \Psi_0(0)| \Psi_0(t_f)\rangle|^{2}= \frac{2 \sqrt{\omega_0 \omega_f}}{\omega_0+\omega_f},
\eeq
and the relation
\beq
\overline{\Delta H_0}\, t_f \geq \hbar \arccos\left[\frac{\sqrt{2} (\omega_0 \omega_f)^{1/4}}{(\omega_0 +\omega_f)^{1/2}}\right].
\eeq
%
%
%
%This implies the lower bound
%
%\beq
%\label{bound A-A}
%\overline{{\Delta E}_0}  \geq \arccos\left[\frac{\sqrt{2} (\omega_0 \omega_f)^{1/4}}{(\omega_0 +\omega_f)^{1/2}}\right] \frac{\hbar}{t_f},
%\eeq
%
%which is evidently a relevant result.
In the $\omega_f \rightarrow 0$ limit
%, in which the right hand side becomes, in dominant order, independent of $\omega_f$.
one finds again the time-energy uncertainty relation for two orthogonal states, that is,
$
\overline{{\Delta H}_0}  \geq  h/ (4 t_f),
$
independent of $\omega_f$.
This bound, although certainly correct, is not tight and
does not capture the actual dependences found
for time averaged standard deviations, which in fact scale on $\omega_f$ and $t_f$ in the same way as the corresponding time averaged energies of the previous section.
Similar to Fig. \ref{fig.2}, Fig. \ref{fig.5} makes the
exponents explicit.
We have used
the standard deviation
$\Delta H_n (t) \equiv [\la \Psi_n|H^2(t)|\Psi_n\ra - E_n^2]^{1/2}$
for the $n$-th expanding mode, which takes the form
\beqa
\label{delta energy}
\Delta H_n(t) &=& \frac{\sqrt{2}  (n^2+n+1)^{1/2} \hbar}{4 \omega_0}
\nonumber\\
&\times&\left[\left(\dot{b}^2+\omega^2(t)b^2+\frac{\omega_0^2}{b^2}\right)^{\!2}
\!\!+\! \frac{4\omega^2_0 \dot{b}^2}{b^2} \right]^{\!1/2}\!\!\!,
\eeqa
and its time average
\beq
\label{average deviation}
\overline{{\Delta H}_n} \equiv\frac{1}{t_f}\int^{t_f}_0 \Delta H_n (t) dt.
\eeq
%

% in $\overline{{\Delta E}_0} \propto t^{\delta}_f %\omega^{\delta}_f $.
The dependence of $\overline{\Delta H_0}$ on $t_f$ and $\omega_f$
as they approach zero independently is summarized by the scaling exponents. In the limit of $t_f \rightarrow 0$,
$\overline{\Delta H_0}\propto t_f^{\delta}$.
Figure \ref{fig.5} shows that $\delta = -2$ for
the calculated standard deviation, whereas $AA$
provides $\delta=-1$.
Similarly, as $\omega_f\to 0$, $\overline{\Delta H_0}\propto \omega_f^{\delta}$.
We find $\delta=-1$ in the calculated standard deviation versus $\delta=0$ from the AA relation.
\section{Adding terms to the Hamiltonian\label{berry}}
Motivated by recent experimental realizations \cite{Nice}, we have  considered up to now simple processes in which the only external manipulation
consists in shaping $\omega(t)$. Other possibilities exist in which the Hamiltonian is complemented with additional terms \cite{Berry09,Muga10}. We shall analyze here the energy excitations
for  a Hamiltonian that results from the transitionless inverse engineering algorythm proposed by Berry \cite{Berry09} (The application to the time dependent oscillator was worked out in \cite{Muga10}.)
We assume here that $\omega(t)$ remains positive,
\beqa
\widetilde{H}&=&H+H_1,
\\
H&=&\hbar \omega(t) (a_t^\dagger a_t+1/2),
\\
H_1&=&i\hbar \frac{\dot{\omega}}{4 \omega}(a_t^2-{a_t^\dagger}^2).
\eeqa
$\widetilde{H}$ would drive the system without transitions
along the states of the instantaneous basis of the time-dependent harmonic oscillator $H$. In particular for the $n$-th state,
\beq
|\phi_n(t)\ra=e^{-\frac{i}{\hbar}\int_0^t \epsilon_n(t') dt'}|n(t)\ra
\eeq
is an exact solution of the time-dependent Schr\"odinger equation with $\widetilde{H}$.

The subscript $t$ in the Schr\"odinger-picture creation and annhilation operators above denotes their fundamental time dependence, because of the changing frequency and eigenstates, not to be confused with the time-dependence of
Heisenberg picture operators. $H_1$ is related to the squeezing operator \cite{Muga10}, and may also be written as $H_1=-\frac{\dot{\omega}}{4\omega}(\hat{q}\hat{p}+\hat{p}\hat{q})$, so that $\widetilde{H}$ is still a generalized harmonic oscillator
quadratic in positions and momenta.
The expectation values of $\widetilde{H}$ and $\widetilde{H}^2$ for the $n$-th state are easily calculated,
\beqa
\la \phi_n|\widetilde{H}|\phi_n\ra&=&\epsilon_n=(n+1/2)\hbar\omega,
\\
\la \phi_n|\widetilde{H}^2|\phi_n\ra&=&
\frac{\hbar^2\dot{\omega}^2}{8 \omega^2}(n^2+n+1)+\epsilon_n^2,
\eeqa
from which we deduce
\beq
\Delta \widetilde{H}_n\equiv(\la\phi_n|\widetilde{H}^2|\phi_n\ra-\epsilon_n^2)^{1/2}
=\frac{\hbar}{4}\frac{|\dot{\omega}|}{\omega}
[2(n^2+n+1)]^{1/2},
\eeq
and the time averages
\beqa
\overline{\epsilon_n}&=&\frac{\hbar(n+1/2)}{t_f}\int_0^{t_f} \omega(t) dt,
\eeqa
\beqa
\overline{\Delta \widetilde{H}_n}&=& \frac{\hbar}{4t_f}[2(n^2+n+1)]^{1/2}\int_0^{t_f} \frac{|\dot{\omega}|}{\omega} dt.
\eeqa
We shall first show with some specific examples
that it is easy to find scalings which are in principle more favourable for implementing a fast transitionless process than the ones in the previous sections.
For example, for a linear frequency ramp expansion $\omega=\omega_0+(\omega_f-\omega_0)t/t_f$,
\beq
\overline{\epsilon_n}=\hbar(n+1/2)\frac{\omega_0+\omega_f}{2},
\eeq
which is independent of $t_f$, and
\beq\label{sdb}
\overline{\Delta \widetilde{H}_n}=\frac{\hbar}{4t_f}[2(n^2+n+1)]^{1/2}\ln\left(\frac{\omega_0}{\omega_f}\right),
\eeq
(in fact a general result for $\dot{\omega}<0$) so, for a fixed $\overline{\Delta \widetilde{H}_n}$, $t_f$ grows only logarithmically as $\omega_f\to 0$.

Note that some of the difficulties with  high transient energies in
the approaches which only control the time dependent frequency, due to the particle exploration of regions far away from the trap center, disappear here since the system evolves at all times along the instantaneous eigenstates without transitions. Clearly
$\la \phi_n|H|\phi_n\ra=\epsilon_n$ and
$\la \phi_n|H^2|\phi_n\ra-\la \phi_n|H|\phi_n\ra^2=0$, so that the standard deviation (\ref{sdb}) is entirely due to the complementary Hamiltonian $H_1$.

The main, and so far important difficulty with this approach is that it is not clear how to implement $\widetilde{H}$ in practice \cite{Muga10}.
$H_1$ involves a non-local interaction and the attempts to provide a quantum-optical realization have not yet succeeded.

\section{Discussion and conclusion}
\label{conclusion}
We have studied the transient energy excitation
in time dependent quantum harmonic oscillators engineered
so that the level populations at a final time are the same as the initial populations.
%Motivated by recent experimental realizations \cite{Nice},
We have  considered first simple processes in which the only external manipulation
consists in shaping $\omega(t)$.  The populations of the instantaneous levels at intermediate times are, however, not preserved, so the transient excitation should be understood and possibly controlled. We have obtained bounds, shown examples, and determined the dominant dependences, which are different from the ones in the Anandan-Aharonov relation \cite{AA}.

In a realistic application the oscillator will not be perfectly harmonic and it is natural to set some maximum
value to the allowed excitation. Then the minimal time required for a fast expansion scales with the final frequency as $t_f\propto
\omega_f^{-1/2}$. As the velocity determining step in  quantum refrigerator Otto cycles
this implies a dependence $R\propto T_c^{3/2}$ of the cooling rate, which had been previously conjectured to be a universal dependence characterizing the unattainability principle for any cooling cycle \cite{Kosloff-EPL}. The present  results provide strong support for the validity of this conjecture
within the set of processes defined exclusively by time-dependent frequencies (without added terms in the Hamiltonian), and call for further testing and study.

In Sect. \ref{berry} we have seen that, at least at a formal level, one could design even faster processes by adding terms to the  harmonic oscillator Hamiltonian, but their physical implementation remains challenging.
\section*{Acknowledgments}
We thank M. Berry, R. Kosloff, A. Ruschhaupt, D. Gu\'ery-Odelin, and E. Torr\'ontegui for  discussions.
Funding by the Basque Government (Grant IT472-10), the
Ministerio de Ciencia e Innovaci\'on (FIS2009-12773-C02-01),
the National Natural Science Foundation of
China (No. 60806041), the Shanghai Rising-Star Program
(No. 08QA14030), the Shanghai Leading Academic Discipline
(No. S30105), and Juan de la Cierva Programme is acknowledged.

\end{document}